# Video-Music Retrieval: A Dual-Path Cross-Modal Network

Xin Gu, Yinghua Shen and Chaohui Lv

*Abstract*—We propose a method to recommend background music for videos. Current work rarely considers the emotional information of music, which is essential for video music retrieval. To achieve this, we design two paths to process content information and emotional information between modal. Based on characteristics of video and music, we design various feature extraction schemes and common representation spaces. More importantly, we propose a way to combine content information with emotional information. Additionally, we make improvements to the classical metric loss to be more suited to this task. Experiments show that this dual path video music retrieval network can effectively merge information. Compare with existing methods, the retrieval task evaluation index: increasing Recall@1 by 3.94 and Recall@25 by 16.36.

## I. Introduction

Video, a form of data that records life and conveys opinions, is produced and uploaded to the Internet by more Internet users. Music as an art form to express emotion is often used to enhance the artistic effect of video. Manually selecting the right music for a video is an extremely lengthy and laborious task. Short video apps often recommend music based on its popularity, and only part of the music conforms to the video. Hence, automatic retrieval of appropriate music clips based on video information is an urgent issue to resolve.

Matching music with a given video belongs to the retrieval task, that is, the corresponding music information is retrieved through the video information. Some researches [1], [2], [3] suggest that both modal have semantic information related to the content. They extract features using a generic network pre-trained on large datasets, which resulted in features contain certain redundant information. Furthermore, due to the diversity of visual content and the richness of musical emotional information, there is some asymmetry between the two modal information. Some researches [4], [5], [6] considers that the emotional information of music is relatively rich and uses the emotional label to restrict the common space. However, they use general networks to pre-train on emotional datasets, so that the network has the ability to learn emotional representations. This method is hard to learn effective emotional information.

In this paper, we combine emotional information and content information to design a dual path cross-modal retrieval network. In the content path, a content common space with fine network structure is designed and additional constraints are added to obtain the content sharing representation of video and music. In the emotional path, the emotional feature extraction scheme is designed to obtain emotional features. The representation of the emotional sharing of video and music is achieved by designing a network structure with simplified emotional common space. Combined with content and emotion, the common fusion space is designed, and the final representation features of video and music are achieved. By computing the similarity between features, the video music retrieval task is finished. Our main contributions are as follows:

1) We construct a video-music retrieval dataset and release the data link.
2) A dual path video-music retrieval network combining content information and emotional information is designed. It can effectively learn various information and use this information to perform retrieval tasks.
3) We design a metric loss function that is more task-consistent.

## II. Related Works

Retrieving music from video is a cross-modal retrieval technology. Unlike other cross-modal retrieval techniques (image-text [7], video-text [8] and video-audio [9]), video music retrieval does not use semantic content information as the only bridge for retrieval. The reason for this difference is mainly the peculiarity of music, which often takes to express emotions as its main content. A positive melody can often inspire the listener, while a negative track can show condolences. Hence, in video-music retrieval, content

The authors are with the College of Information and Communication Engineering, Communication University of China, Beijing 100024, China (e-mail: guxin@cuc.edu.cn; shenwan@cuc.edu.cn; llvch@cuc.edu.cn).
 (the PDF file is attached to the Supporting Documents). Code is available at https://github.com/GuXin34/DPVM. Digital Object Identifier XXX.

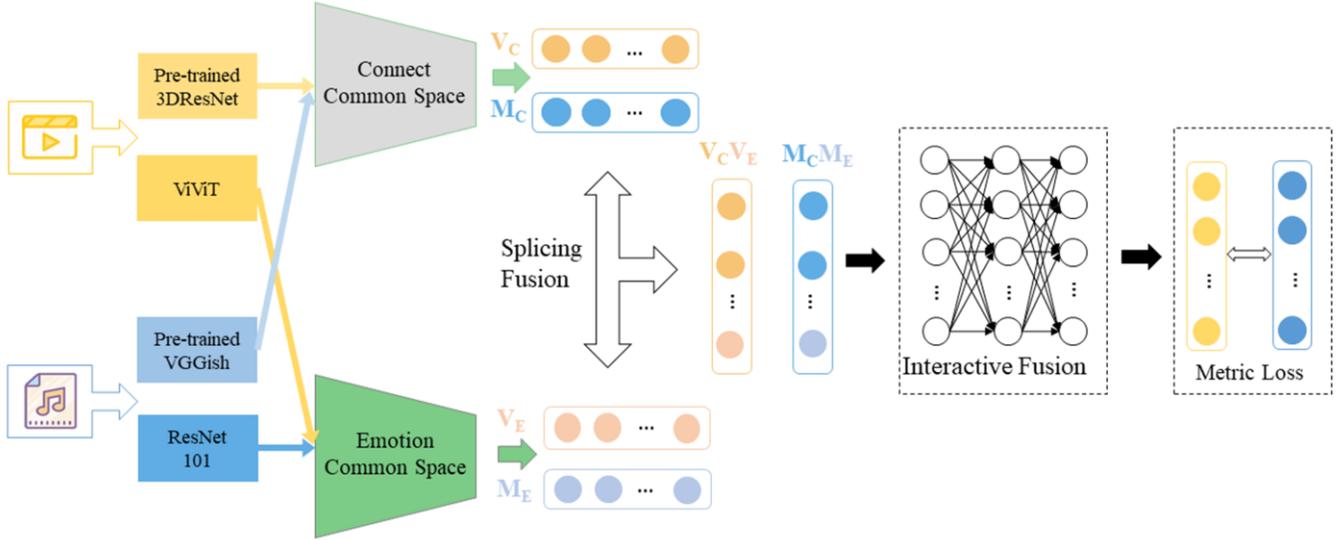

Fig. 1. Overview of the proposed DPVM that consists of a content common network, an emotion common network, and a fusion network. Content features get their shared representations through the content common space composed of encoder-decoder. Emotional features obtain their shared representations through the emotion common space composed of MLP. Splicing Fusion refers to splicing content features and emotional features of the same modal. Interactive Fusion fully interacts with content information and emotional information.

information and emotional information from both modal are crucial.

Content-based video-music retrieval considers that both modalities have semantic content-related information, like the consistency of the events of the video and the theme of the music, etc. Consequently, they tend to be self-supervised methods, using paired video and music pairs without additional label information. Hong S et al. [1] propose the Cbvmr model, which uses traditional machine learning methods to extract the content features, and uses triple loss to limit the formation of the common representation space. Based on the Cbvmr model, Pretel L et al. [2] use a deep learning network model to replace the method of manually extracting features. Yi J et al. [10] propose a CMVAE model using textual information. Pretet L et al. [3] propose and verify the effectiveness of movement information. The model uses Variational AutoEncoders(VAE) [11] as a common representation space and is optimized by generative loss functions. Zhang et al. [12] design a VAE-CCA network combining the encoder-decoder structure. The network utilizes the decoder to reconstruct the features from the encoder, ensuring that the features reflect the structure of the original data, thus improving retrieval performance. Surís et al. [13] gave more attention to contextual long-term information. They use CLIP [14] to extract visual information from video frames, use Inception [15] to extract musical information from music segments. Lastly, the sequence information is entered into the common representation space structured by Transformer [16] to get the common features.

Emotion-based video-music retrieval considers that music is often used as media information to express emotions, so the emotional information of music is relatively rich. Aiming for that characteristic of music, research often use emotional information to restrict the common representation space. Shin K et al. [4] use linear regression to predict the emotion of video and music, and compute the Euclidean distance between video and music data. Using emotional labels, Zeng D et al. [5] propose a supervised DCCA (S-DCCA) model as a common representation space. Li B et al. [6] train video networks and music networks separately on emotion datasets, making those networks to learn emotional information. Shang L et al. [17] believed that the connotative association of images and music included lyrical emotions in addition to semantic concepts, so they developed a connotation-aware music retrieval framework (CaMR).

### III. METHOD

Fig. 1 shows our proposed dual-path cross-modal retrieval scheme DPVM (Dual Path Video Music retrieval network) that consists of a content common network, an emotion common network, and a fusion network.

#### A. The Content Common Network

As shown in Fig. 2, for video, we use the pre-trained 3DResNet [18] on the Kinetics-600 dataset [19] to extract its content features. For music, we use the pre-trained Vggish [20] on the AudioSet to extract its content features. In the design of the content common space, an encoder-decoder structure is adopted. The encoder reduces the dimension, and unifies the dimension of video and music features while sharing weights. The decoder is used to reconstruct the dimension. At the same time, it ensures that important information about dimensionality reduction features is not lost.

There are two parts which compose the optimization loss for content common networks: reconstruction loss and metric loss. Reconstruction loss optimizes encoder dimensionality reduction processing for content features. Metric loss optimizes the consistency of common feature information for generated video and music content.

Given video feature $V_i$ and music feature $M_i$. The reconstruction loss:

$$L_R = \| V_i - D(E(V_i)) \| + \| M_i - D(E(M_i)) \|. \tag{1}$$

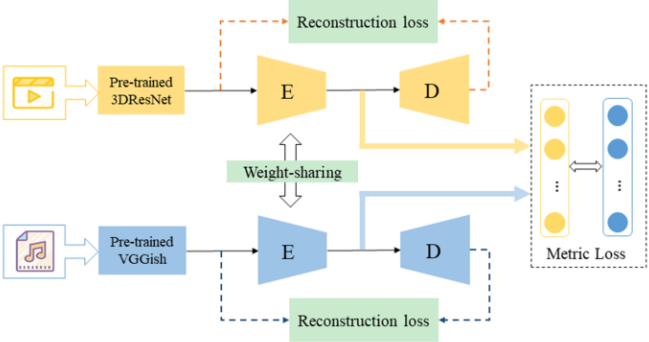

Fig. 2. Overview of the proposed content common network. E denotes encoder structure, D denotes decoder structure.

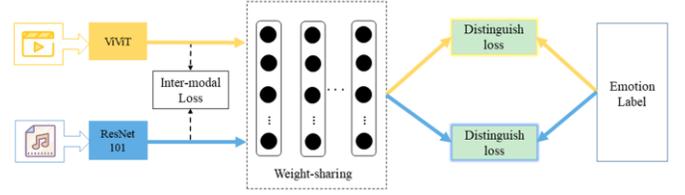

Fig. 3. Overview of the proposed emotion common network. MLP as emotional common representation space.

The metric loss:

$$L_M = \frac{1}{2}(1-Y)\parallel E(V_i) - E(M_i)\parallel^2 + \frac{1}{2}(Y)\{max(0, margin - \parallel E(V_i) - E(M_i)\parallel)\}^2. \quad (2)$$

The content loss:

$$L_{Content} = \lambda_1 L_R + \lambda_2 L_M. \quad (3)$$

Where $\parallel \cdot \parallel$ denotes the cosine distance, $E(\cdot)$ denotes the encoder, $D(\cdot)$ denotes the decoder. Y is paired labels, if $\{V_i, M_i\}$ is matching data, then Y=0, otherwise Y=1. The margin represents the distance threshold, $\lambda_1$ and $\lambda_2$ represent weight coefficients.

### B. The Emotion Common Network

As shown in Fig. 3, because the emotional content of the video is sparse [21], that is to say the information of the video frame is redundant, and only a few frames contain rich emotional information. To efficiently get the emotional information from the video, we first choose the emotional key frames [22] from the video. Second, in the feature extraction network, the ViViT [23] visual transformer model is selected. This not only improve the running speed but also obtain the temporal context information of the video. In order to effectively obtain emotional information from music, the ResNet101 [24] is adopted and the channel attention mechanism is introduced. Design an emotional common space with MLP as the structure.

There are two parts which compose the optimization loss for emotion common networks: discrimination loss and inter-modal loss. The discrimination loss adopts the cross-entropy loss function.

$$L_D = -\sum Y(V_i) \log \frac{\exp(F(V_i))}{\sum_{i=1}^n \exp(F(V_i))} - \sum Y(M_i) \log \frac{\exp(F(M_i))}{\sum_{i=1}^n \exp(F(M_i))}. \quad (4)$$

Inter-modal loss, preserves the difference in sentiment characteristics of various modal data. This benefits the emotional common space to form pure emotional discrimination information from video and music.

$$L_M = -\parallel V_i - M_i \parallel. \quad (5)$$

The emotion loss:

$$L_{Emotion} = \mu_1 L_D + \mu_2 L_M. \quad (6)$$

Where $Y(\cdot)$ is the real emotion label, $F(\cdot)$ denotes the emotion common representation space of the multilayer perceptron structure, and $\mu_1$ and $\mu_2$ represent the weight coefficients.

### C. The Fusion Common Network

The fusion features are obtained by splicing two different features of the same modal data. This fusion method ignores the information connection between content and emotion to some certain extent. Therefore, we design a common fusion space based on splicing fusion. The splicing features obtain interactive fusion features by fusing the common space. This common space with the fully connected layer (FC) that can make content information and emotional information interact.

The loss function optimized for common fusion space is the loss of contrast between interactive fusion features:

$$L_{Fusion} = \parallel F(V) - F(M) \parallel. \quad (7)$$

Combining equations (3), (6) and (7), we obtain the objective function of the proposed method DPVM as:

$$\text{Loss} = k_1 L_{Content} + k_2 L_{Emotion} + k_3 L_{Fusion}. \quad (8)$$

### D. The Polarity Penalty Metric Loss

In metric loss, the model is optimized by increasing the similarity between related data pairs and reducing the similarity between unrelated data pairs [25]. To some extent, there are some unrelated constraints and low relevance. In a batch with N pairs of data, $\{v_k, m_k\}$ represents paired video-music data, $\varphi(v_k, m_i)$ is the similarity between the data.

The metric loss:

$$L_{\varphi(v,m)} = \sum_k^N \sum_{i \neq k}^N (\varphi(v_k, m_i) - \varphi(v_k, m_k)). \quad (9)$$

If $\varphi(v_k, m_j) < \varphi(v_k, m_k)$, then $v_k$ and $m_j$ are called low-similarity data pairs. Repeatedly processing these low-similarity data pairs will cause the optimization function to focus on irrelevant targets for many times, resulting in low optimization efficiency. Therefore, we propose a more efficient approach based on the existing loss function.

Firstly, with regard to the issue of redundancy of the metric weight loss coefficient, the metric loss coefficient $\rho_{ki}$ of the low similarity data pair is set to 0.

$$\rho_{ki} = \begin{cases} 0, & \varphi(v_k, m_i) < \varphi(v_k, m_k) \\ 1, & otherwie \end{cases}. \quad (10)$$

Given the low weight coefficient and task fitness, and the issue of label difference is not taken into account, a polarity penalty coefficient $P_{ki}$ is proposed.

$$P_{ki} = |label_k - label_i|. \quad (11)$$

The polarity penalty metric loss as:

$$L_{\varphi(v,m)} = \sum_k^N \sum_{i \neq k}^N (P_{ki} \rho_{ki} \varphi(v_k, m_i) - \varphi(v_k, m_k)). \quad (12)$$

TABEL I

VIDEO TO MUSIC RETRIEVAL RESULTS ON MVED

| Method | Recall @1 ↑ | Recall @5 ↑ | Recall @10 ↑ | Recall @15 ↑ | Recall @20 ↑ | Recall @25 ↑ |
|---|---|---|---|---|---|---|
| Baseline-Emotion [6] | 5.39 | 10.61 | 15.37 | 20.08 | 23.53 | 26.17 |
| Baseline-Connect [2] | 7.59 | 15.23 | 20.31 | 26.25 | 30.13 | 34.27 |
| $DPVM_E$ | 6.43 | 13.54 | 20.26 | 25.09 | 31.41 | 37.26 |
| $DPVM_C$ | 8.31 | 16.42 | 21.21 | 27.99 | 33.94 | 39.01 |
| $DPVM_{Sp}$ | 9.13 | 16.94 | 22.33 | 29.50 | 37.41 | 42.35 |
| $DPVM_{In}$ | **10.94** | **18.84** | **24.39** | **34.32** | **43.26** | **49.97** |

TABEL II

VIDEO TO MUSIC RETRIEVAL RESULTS WITH PPML ON MVED

| Method | Recall @1 ↑ | Recall @5 ↑ | Recall @10 ↑ | Recall @15 ↑ | Recall @20 ↑ | Recall @25 ↑ |
|---|---|---|---|---|---|---|
| Baseline-Emotion [6] | 5.64 | 11.25 | 16.80 | 21.58 | 24.93 | 27.33 |
| Baseline-Connect [2] | 7.91 | 16.43 | 22.92 | 27.53 | 31.29 | 35.90 |
| $DPVM_E$ | 7.17 | 14.85 | 21.44 | 26.73 | 32.05 | 38.16 |
| $DPVM_C$ | 9.29 | 17.86 | 23.21 | 28.65 | 34.30 | 39.47 |
| $DPVM_{Sp}$ | 10.42 | 18.34 | 24.07 | 31.74 | 38.26 | 43.19 |
| $DPVM_{In}$ | **11.53** | **20.11** | **26.67** | **36.01** | **44.87** | **50.63** |

## IV. EXPERIMENTS

### A. Dataset

Music compositions are protected by copyright, making dataset difficult to publish. Hong S et al. [1] obtained the HIMV-200K benchmark dataset from the YouTube-8M large-scale labelled video dataset [26]. Due to the video link failure issue, the data available is 50K [2] in 2021 and 20K [13] in 2022.

For this purpose, we select well-known cinematic works from the film platform and capture video clips that contain only background music. We modify these clips to ensure that the duration of the video is about 15 seconds, forming a MVED dataset of about 3k video-music pairs. Emotional descriptions like 'sad, happy, scared, surprised' and other adjectives were also provided. By combining these emotional descriptors, the dataset provides three polarity labels: 'positive, negative, neutral'. According to our research, 5 pieces of data-related music are selected for each video. The final MVED contains 10k video-music pairs, which are split into a training set and a test set based on the 7:3 ratio.

### B. Implementation Details

For video, the redundant frames are removed by taking 6 frames at equal intervals within 1 second. In the emotion path, the emotional key frames are 16 frames. For music, MFCC features are extracted by Librosa library in Python. MFCC has the time-frequency domain information of music, and has the characteristics of high recognition ability and anti-noise [27].

We use Adam optimizer [28] with {β1 = 0.5, β2 = 0.999} and train the model for 100 epochs. The learning rate is set to 1e−4, and the batch size is 16. Metric loss uses the calculation method of cosine distance. In (3), (6) and (8), $\lambda_1$=0.8、$\lambda_2$=1.0; $\mu_1 = 0.8$、$\mu_2 = 1.0$; $k_1 = 0.5$、$k_2 = 0.5$、$k_3 = 1.0$.

### C. Results

To evaluate our method, we compare its performance with methods from related work. Table I shows the experimental results on the MEVD dataset.

The benchmark network is the prior method of Prétet et al. [2], MLP is used as the common space and optimized with triplet loss. They use ImageNet to extract video features and utilise OpenL3 [29] to extract music features. The dual stream emotional network [6] only pre-trained the network on the emotional dataset, and does not further extract emotional information. In our work, these two classic networks are called content-based benchmark networks and emotion-based benchmark networks. Among them, $DPVM_C$ refers to the single content path network, $DPVM_E$ refers to a single emotional path network. $DPVM_{Sp}$ is DPVM using splicing fusion strategy, $DPVM_{In}$ is DPVM using interactive fusion strategy.

Based on the above experiments, we replace the metric loss with the polarity penalty metric loss(PPML), and the results are shown in Table II.

The results indicate that the content information path is more efficient than the emotion information path. This is because feature extraction networks are pre-trained on large-scale dataset. However, there currently exists a lack of such pre-trained networks for emotional information. The DPVM network using interactive fusion strategy works best. Moreover, the suggested metric loss of polarity penalty also shows good performance.

## V. CONCLUSION

In video music retrieval work, we emphasize the importance of emotional information. By designing a dual path network to integrate content information with emotional information. Because of the difference between content information and emotional information, we have various designs for feature extraction and common representation space. Specifically, the content information is extracted through the pre-trained network, and the content common space of the coding and decoding structure is designed to remove redundant information. The path of emotional information adopts the emotional scheme, and constructs the common space of emotional with the MLP as a structure.

We also further explored ways to integrate content with emotion. Furthermore, we optimize the classical metric loss function according to the characteristics of the task. As part of future work, we will continue to conduct in-depth research on the interaction between emotion and content.